\pdfoutput=1 % only if pdf/png/jpg images are used
\documentclass{JINST}

\title{The NIFFTE project}

\author{J. Ruz$^a$\thanks{Corresponding author.}~,
D.M Asner$^b$, R.G Baker$^c$, J. Bundgaard$^d$, E. Burgett$^e$, M. Cunningham$^a$, J. Deaven$^e$, D.L. Duke$^d$, U. Greife$^d$,S. Grimes$^f$, M. Heffner$^a$,
T. Hill$^g$, D. Isenhower$^h$, J.L. Klay$^c$, V. Kleinrath$^e$, N. Kornilov$^f$, A.B. Laptev$^i$, W. Loveland$^j$, T.N. Massey$^f$, R. Meharchand$^i$, 
H. Qu$^h$, S. Sangiorgio$^a$, B. Seilhan$^a$, L. Snyder$^a$, S. Stave$^b$, G. Tatishvili$^b$, R.T. Thornton$^h$, F. Tovesson$^i$, D. Towell$^h$, 
R.S. Towell$^h$, S. Watson$^h$, B. Wendt$^e$  
and L. Wood$^b$\\
\llap{$^a$}Lawrence Livermore National Laboratory, Livermore, CA 94550, USA\\
\llap{$^b$}Pacific Northwest National Laboratory, Richland, WA 99354, USA\\
\llap{$^c$}California Polytechnic State University, San Luis Obispo, CA 93407, USA\\
\llap{$^d$}Colorado School of Mines, Golden, CO 80401, USA\\
\llap{$^e$}Idaho State University, Pocatello, ID 83209, USA\\
\llap{$^f$}Ohio University, Athens, OH 45701, USA\\
\llap{$^g$}Idaho National Laboratory, Idaho Falls, ID 83415, USA\\
\llap{$^h$}Abilene Christian University, Abilene, TX 79699, USA\\
\llap{$^i$}Los Alamos National Laboratory, Los Alamos, NM 87545, USA\\
\llap{$^j$}Oregon State University, Corvallis, OR 97331, USA\\
E-mail: \email{ruzarmendari1@llnl.gov}}

\abstract{The Neutron Induced Fission Fragment Tracking Experiment (NIFFTE) is a double-sided Time Projection Chamber (TPC) with micromegas 
readout designed to measure the energy-dependent neutron-induced fission cross sections of the major and minor actinides with unprecedented 
accuracy. The NIFFTE project addresses the challenge of minimizing major sources of systematic uncertainties from previous fission chamber 
measurements such as: target and beam non-uniformities, misidentification of alpha and light charged particles as fission fragments, and 
uncertainties inherent to the reference standards used. In-beam tests of the NIFFTE TPC at the Los Alamos Neutron Science Center (LANSCE) 
started in 2010 and have continued in 2011, 2012 and 2013. An overview of the NIFFTE TPC status and performance at LANSCE will be presented.}

\keywords{Fission; TPC; Cross Section}

\begin{document}

\section{Introduction}\label{sec:intro}

Sensitivity studies indicate that high-accuracy measurements of fission cross sections are needed  and the Neutron Induced Fission Fragment Tracking 
Experiment (NIFFTE) collaboration is developing a Time Projection Chamber (TPC) to meet this need. Neutron-induced fission cross sections have been 
measured with fission chambers (ICs) for decades, typically by placing a thin target of the fissile material of interest into a neutron beam and 
detecting the ionization left in a gas by the resulting fission products. To date, fission chambers have been limited to measuring the energy 
deposited in a gas by charged particles, and were therefore unable to distinguish between different particles of the same energy. Fragments are 
subject to energy loss and straggling within the target\footnote{A non negligible fraction of fission fragments might become indistinguishable 
from lighter charged particles within the fission chamber due to energy loss and straggling.} resulting in a systematic error for the measured 
cross section. 
 
Based on well-established technology originally developed for high-energy physics experiments, the NIFFTE TPC is uniquely suited to achieve 
unprecedented accuracy. The detection principle of a TPC is rather simple: particles ionizing the gas traverse a uniform electric 
field. The electrons produced by the ionization are drifted by the electric field and collected at a segmented readout plane. A two-dimensional 
projection of the particle's track is obtained by interpreting the amount of charge deposited in each segment. Combining this two dimensional projection 
with the drift time of the electrons provides a full, three-dimensional representation of the particle's path through the detector.

\begin{figure}[tbp]   
\centering
\begin{center}
\includegraphics[width=.44\textwidth,height=0.35\textwidth]{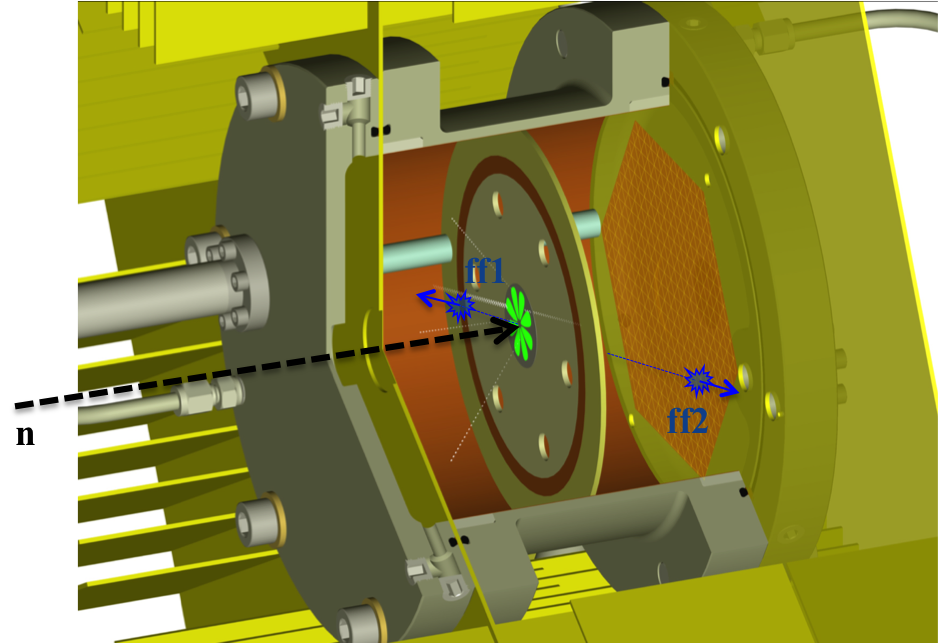}
\hspace{1.25 cm}
\includegraphics[width=.415\textwidth,height=0.35\textwidth]{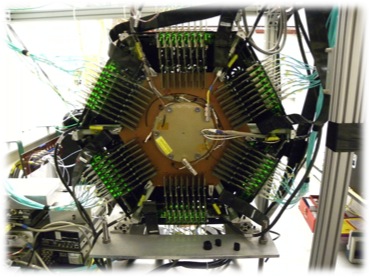}
\end{center}
\caption{Schematic of a neutron-induced fission in the target of the NIFFTE TPC (left) and picture of the TPC installed in the LANSCE-WNR facility 
during the 2012 run cycle.}
\label{fig:sch}
\end{figure}

A schematic of a neutron-induced fission in the TPC is shown in figure \ref{fig:sch}. The fissionable sample to be studied is placed in the center 
of the detector at the midpoint of the cathode. The field cage covers the $\rm{54\,mm}$ drift distance at each side of the sample and it is placed 
inside a pressure vessel of $\rm{15\,cm}$ diameter that is filled up with a suitable gas mixture e.g. 
Argon($90\,\%$)-Methane($10\,\%$) at a nominal $\rm{550\,torr}$. At each end of the chamber charge is amplified and read out on a MICROMEGAS ``pad'' 
plane. The signals are digitized and processed by custom-made EtherDAQ card assemblies \cite{DAQpaper,r1} sampling at $50\,\rm{MHz}$ together with a 
software framework developed for the data acquisition, analysis, and simulation tasks \cite{r2}. An article describing the TPC hardware is in preparation.

\section{Beam experiments}\label{sec:beam}
Building a TPC suitable for fission research is an ambitious task, involving complex systems and state-of-the-art technology. As new hardware and 
software capabilities are implemented, frequent evaluation is required. In-beam testing takes place at the 4FP90L flight path of Los Alamos Neutron 
Science Center (LANSCE) Weapons Neutron Research (WNR) facility \cite{r3} that makes use of an unmoderate tungsten spallation target to produce 
a high-energy neutron spectrum with energies ranging from $100\,\rm{keV}$ to $600\,\rm{MeV}$.  

The first TPC experiments at LANSCE-WNR took place in 2010. Data were collected first with 64 then 192 individual channels ($\rm{1\,\%}$ and 
$\rm{3\,\%}$ of the full TPC), with blank carbon samples and on carbon backings with $\rm{^{238}U}$ deposited onto them \cite{r4}. These 
early measurements validated the TPC technique, as particle tracks could be reconstructed for both fission fragments and alpha particles. During 
the 2011 LANSCE run cycle, one complete sextant (496 channels, $\rm{8\,\%}$ of the full TPC) was instrumented, and data were collected with samples 
containing $\rm{^{238}U}$, $\rm{^{235}U}$, and $\rm{^{239}Pu}$. During the most recent run cycle (2012), the TPC was instrumented with one complete 
pad plane (2976 channels, 1/2 of the full detector), and notably, for the first time in the TPC project, information about the neutron time-of-flight 
was available.

\section{Performance}\label{sec:perf}
Preliminary analysis of the data taken during the 2010 and 2011 LANSCE run cycles have been reported \cite{r5,r6} and the 2012 run cycle data is 
currently being analyzed. Data were collected on three samples: a split sample (semicircular deposits) containing $\rm{^{238}U}$ and $\rm{^{235}U}$; a split 
sample with $\rm{^{235}U}$ and $\rm{^{239}Pu}$; and a sample with $\rm{^{239}Pu}$ electro-plated onto metallized polyproplylene. In adition, off-beam data 
was collected with the use of $\rm{^{252}Cf}$ and $\rm{^{244}Cm}$ bottom sources to fine tune the acquisition software and the analysis codes.  
All results included in this paper correspond to the 2012 data taking period and should be considered preliminary. 

\subsection{Neutron Time of Flight}\label{sec:ntof}
\begin{figure}[tbph]
\centering
\includegraphics[width=.7\textwidth]{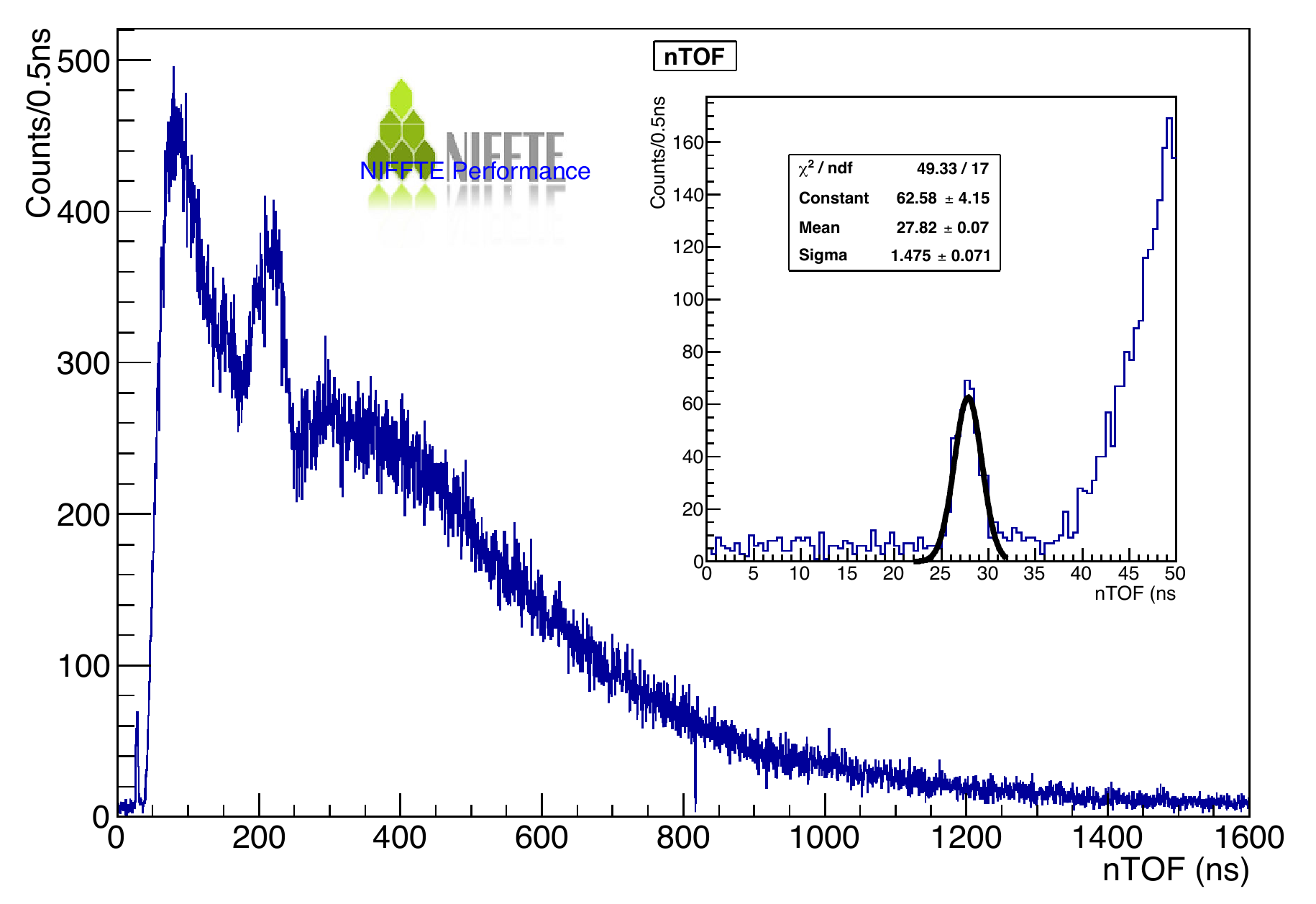}%{NTOFDNP.pdf}%{NTOF_C1.pdf}
\caption{Neutron time of flight reconstruction for neutron-induced fission events in the TPC and zoom of the observed photo-fission peak. A split sample 
(semicircle deposits) containing $\rm{^{238}U}$ and $\rm{^{235}U}$ was used during this measurement.}
\label{fig:ntof}
\end{figure}

Neutron-induced fission depends on the energy of the incident neutron. The neutron Time-Of-Flight (nTOF) between the spallation target at WNR and the 
fissionable target placed inside the TPC can be folded together with the overall WNR beam spacing structure -nominal $1.8\,\rm{\mu s}$ pulse spacing 
between neutron bunches -to reproduce the energy spectrum of the neutrons inducing fission events in the TPC. Figure \ref{fig:ntof} 
shows a reconstruction of the nTOF for the NIFFTE TPC during the 2012 run cycle. The peak around $28\,\rm{nsec}$ corresponds to photo-induced 
fission in the TPC target and, can be used to obtain an absolute calibration for both, the distance of the TPC relative to the spallation target 
and the energy of the incoming neutrons. The width of the photo-fission peak serves as a tool to determine the systematic error on the energy 
of the incident neutron. During 2012, the NIFFTE TPC was able to achieve $2.5-3.5\,\rm{nsec}$ FWHM for the photo-fission peak, which is competitive 
with the resolution obtained by previous fission chamber experiments.

\subsection{Tracking and Particle Identification}\label{sec:track}
One of the major advantages of using a TPC to study neutron-induced fission cross sections is the ability to reconstruct and study the features of the 
ionizing tracks that the different particles generate on their way through the active volume of NIFFTE. 

\begin{figure}[tbph] 
\centering
\includegraphics[width=.7\textwidth]{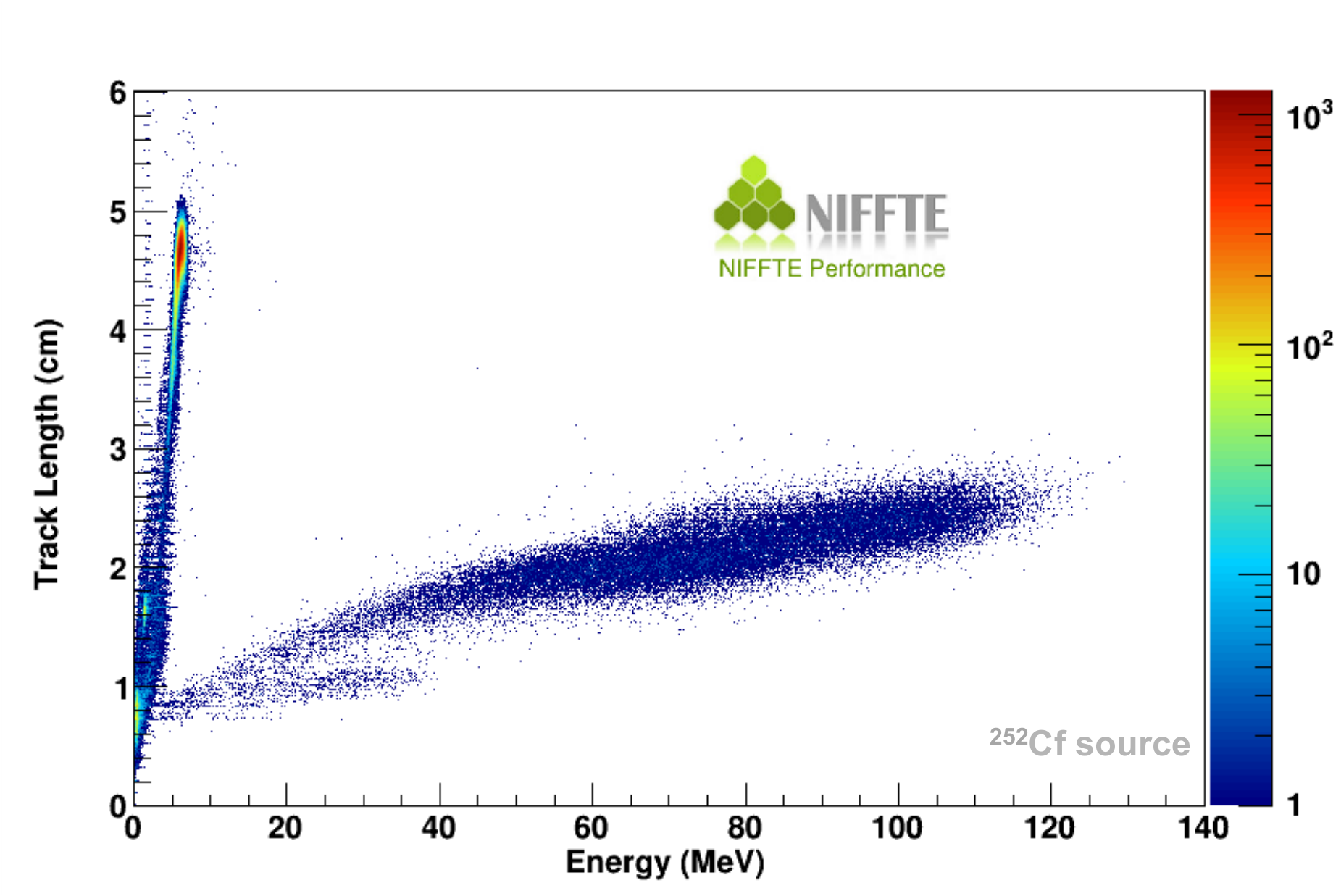}
\caption{Plot of length versus energy of track reconstructions in the TPC for a $\rm{^{252}Cf}$ buttom source. On the plot there are three regions 
that can be easily spotted, an almost vertical line at low energies with lengths from $\rm{1\,cm}$ up to $\rm{6\,cm}$ containing $\alpha$-particles and protons,
a big elliptical zone centered around $\rm{80\,MeV}$ with tracks from  $\rm{1.5\,cm}$ up to $\rm{3\,cm}$ gathering most of the fission 
fragment population and a small shadow to the elipse with track lengths around  $\rm{1\,cm}$ and energies between $\rm{5}$ and $\rm{40\,MeV}$. The 
shadow region has been succesfully identified as fission fragments that come out of the target at high polar angles, suffering from straggling, and 
hit on the concentric copper ring of $\rm{2}$ by $\rm{4\,cm}$ diameter that holds the polypropylene backing foil where the $\rm{2\,mm}$ diameter 
$\rm{^{252}Cf}$ spot was deposited.}
\label{fig:LE}
\end{figure}

One example of how particle identification (distinguishing alpha and light charged particles from fission fragments) might be done using the tracking 
information of the TPC can be shown by comparing the length of the particle's track to its total energy deposited in the TPC 
(see figure \ref{fig:LE}). Fission fragments, have relatively short tracks and deposit large amount of energy in the fill gas, while light particles 
(alpha particles) have longer tracks and deposit less energy. 

\begin{figure}[tbph] 
\centering
%\includegraphics[width=.48\textwidth]{Alpha2.pdf}
%\hspace{0.3 cm}
\includegraphics[width=.6\textwidth]{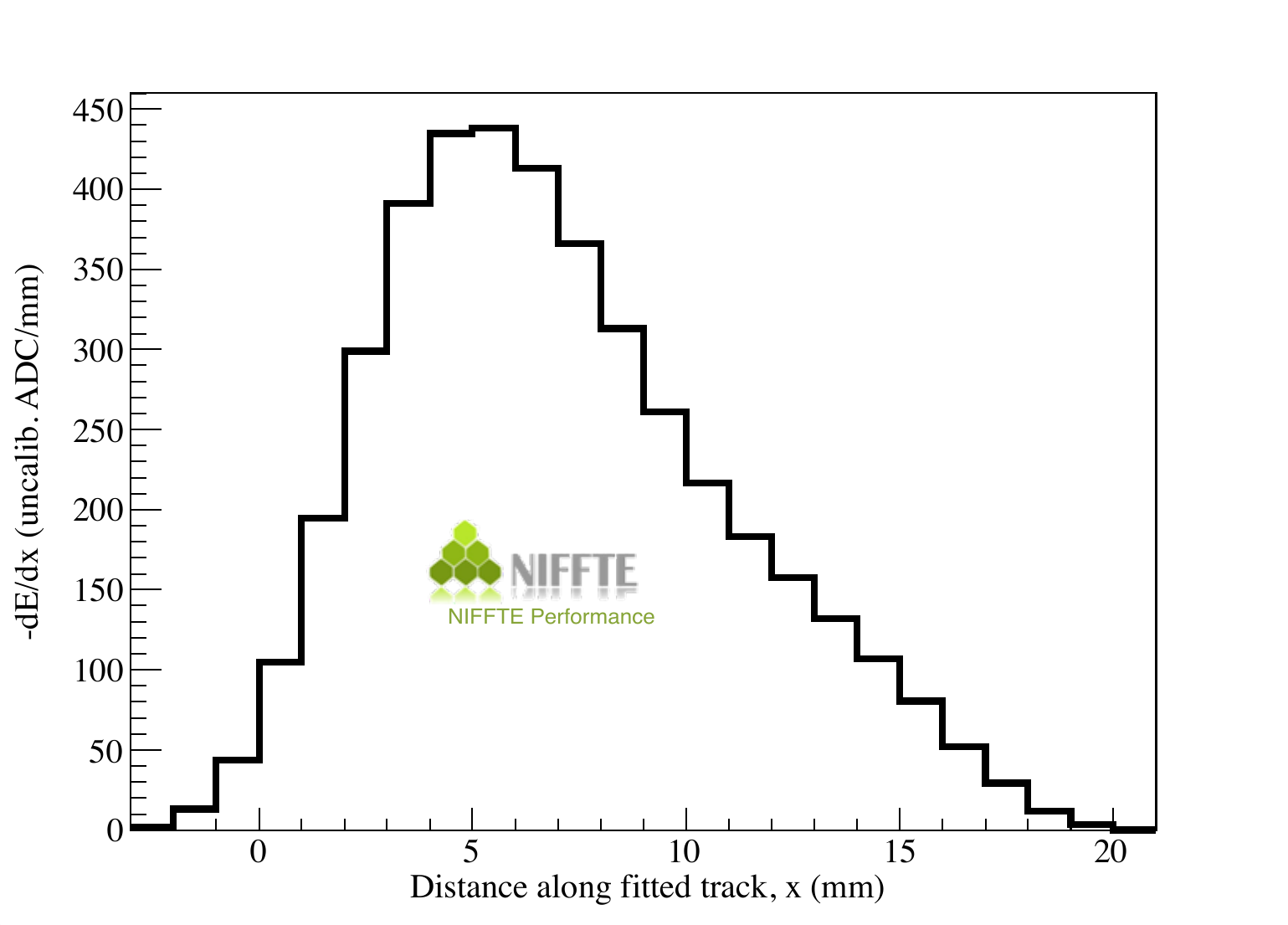}
\caption{Ionization pattern of a fission fragment in the NIFFTE TPC. The Bragg peak is located at the begining of the track for these particles.}
%\caption{Ionization pattern of an $\alpha$-particle (left) and a fission fragment (right) in the NIFFTE TPC. On the left it can also be noticed a 
%comparison between the observed ionization track (black) and a theoretical prediction model that  includes the ionization nature of an 
%$\alpha$-like particle together with the specific diffusion expected in the TPC for that particular track (blue). In red the residual between the observed
%and modeled ionization pattern.}
\label{fig:PID1}
\end{figure}

\begin{figure}[tbph] 
\centering
\includegraphics[width=.6\textwidth]{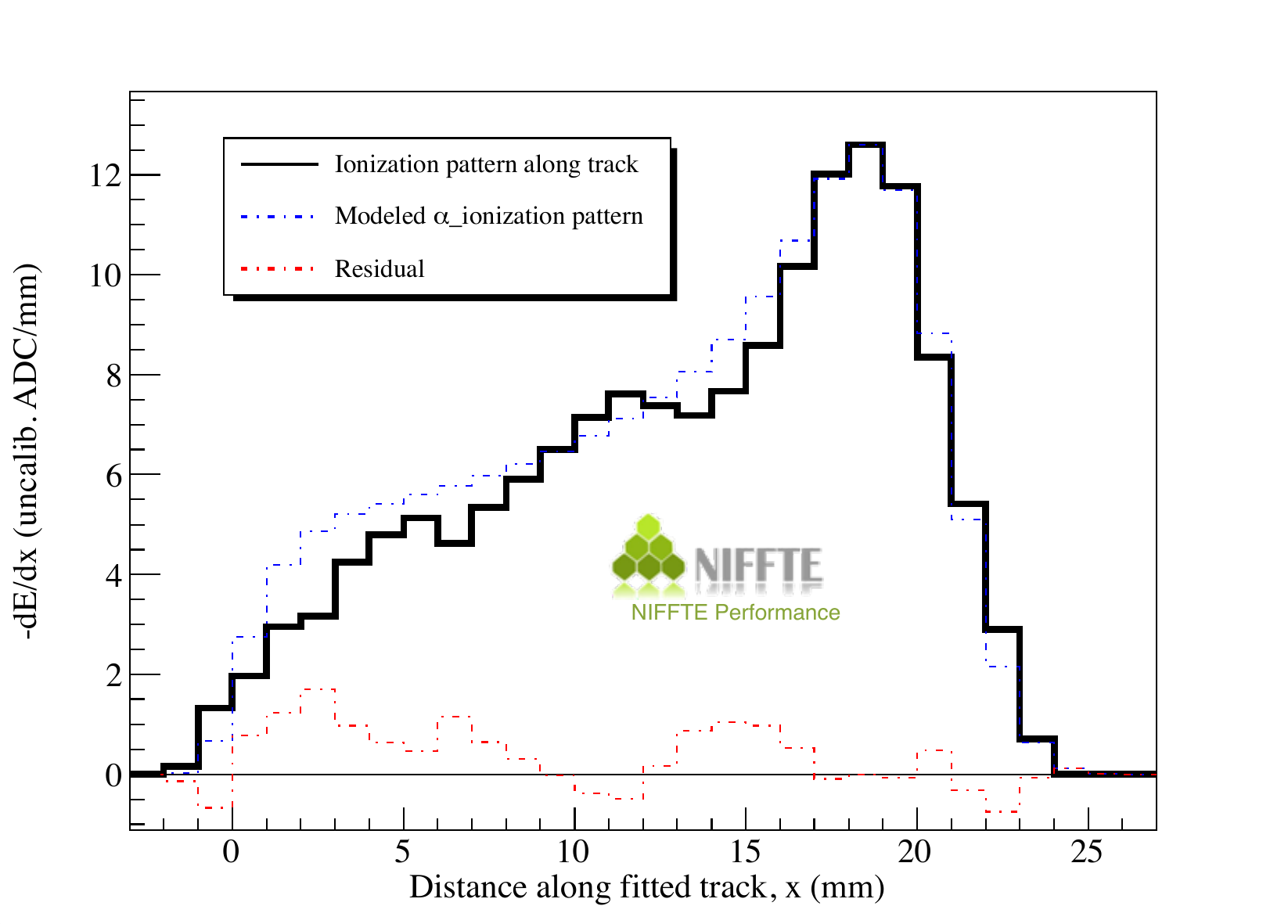}
\caption{Ionization pattern of an observed $\alpha$-track in the NIFFTE TPC (black) and a theoretical prediction that  includes 
the ionization nature of an $\alpha$-like particle together with the specific diffusion expected in the TPC for this particular track (blue). 
In red the residual of observed and modeled ionization pattern is shown.}
\label{fig:PID2}
\end{figure}

Using precise tracking information, e.g. the differential ionization pattern of the particle within the gas, allows the search of the specific Bragg 
peaks created by the particles and the comparison of their ionization pattern with simulations and theoretical models to provide an ultimate tool 
for particle identification. Fission fragments deposit most of the energy at the begining of their short track through the TPC, while $\alpha$ particles 
and protons penetrate the gas further and have their Bragg peak towards the end of their path through the TPC (see figures \ref{fig:PID1} and 
\ref{fig:PID2}).

Finally, tracking information also enables tracing of each individual particle back to a well-defined origin in the sample, which ultimately allows for 
a minimization on the systematics errors such as beam spot size, and sample uniformity. This makes it possible to study two or more actinides 
on the same backing, as described in \cite{r4}, and to check sample uniformity using the decay activity of the actinide deposits. 

\begin{figure}[tbph]
\centering
\includegraphics[width=.6\textwidth]{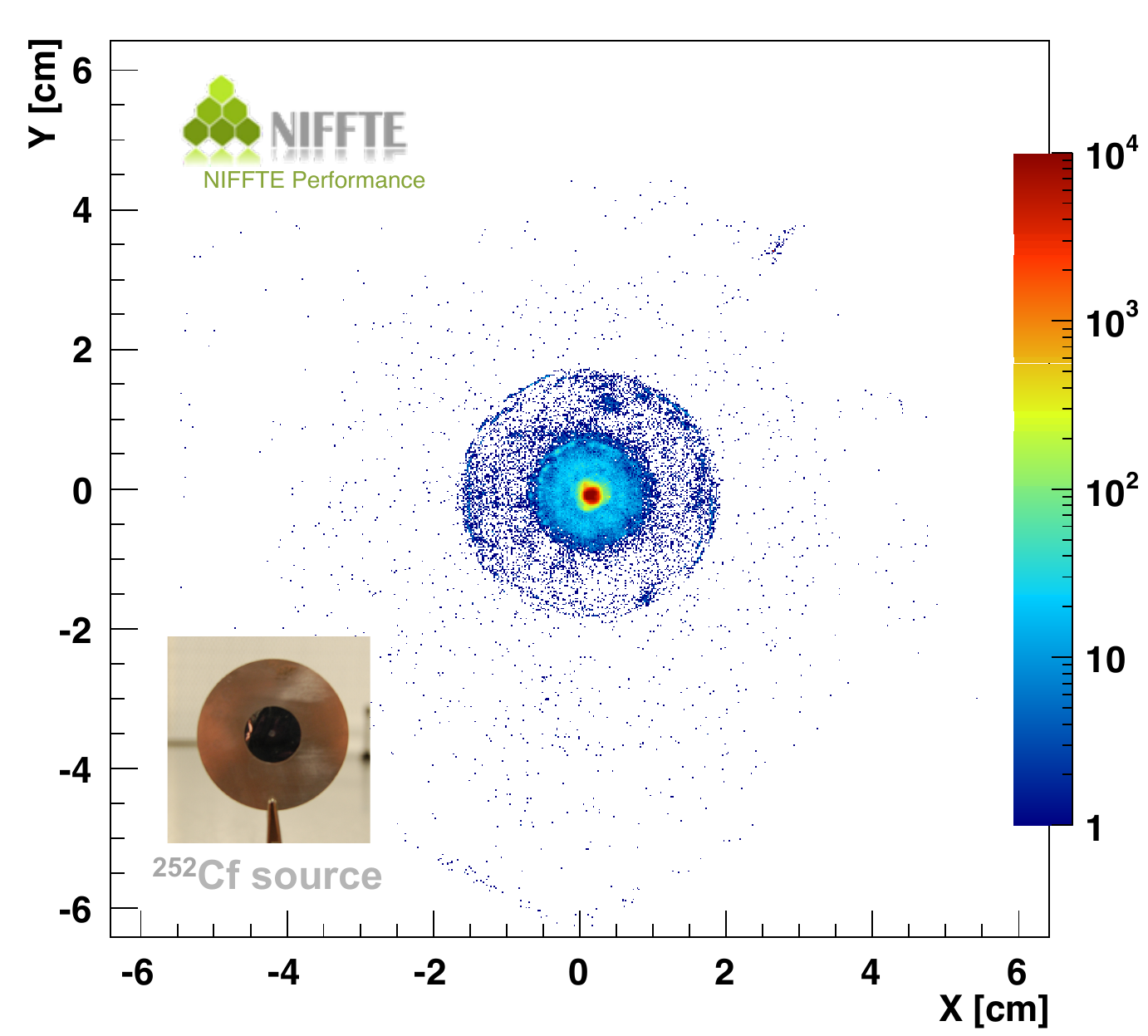}%{Cf-Vertex2.pdf}
\caption{Track vertex reconstructions of $\alpha$-like events emerging from a $\rm{2\,mm}$ spot source of $\rm{^{252}Cf}$ placed in the center 
of the NIFFTE TPC. In the figure a picture of the sample used can be seen.}
\label{fig:vertex}
\end{figure}

Figure \ref{fig:vertex} shows the vertex reconstructions for spontaneous $\alpha$-like events from a $\rm{^{252}Cf}$ bottom source. The tracks 
seem to originate from the center of the cathode plane, coming out of a circle of approximately $\rm{2\,mm}$ diameter, which is consistent with 
the actual dimension of the $\rm{^{252}Cf}$ deposit. Note that the two different backing materials can easily be distinguished: one external 
ring of copper with inner and outer diameters of $\rm{2\,cm}$ and $\rm{4\,cm}$ respectively, and a  $\rm{2\,cm}$ diameter polypropylene foil on top of 
which the $\rm{^{252}Cf}$ was deposited. This figure clearly shows the radiographic capabilities of NIFFTE to map sample uniformities.

\section{Conclusion}
The NIFFTE TPC is being tested off-beam and in-beam at the neutron facility LANSCE-WNR. Preliminary results (with half detector instrumented) 
indicate that the TPC performance is meeting expectations. As the NIFFTE project moves from development into production phase, the collaboration is 
optmistic that the project is on-track to deliver neutron-induced fission cross sections with unprecedented accuracy.

\acknowledgments
This work was performed under the auspices of the U.S. Department of Energy by Lawrence Livermore National Laboratory under contract 
No. DE-AC52-07NA27344.

\end{document}